\documentclass[aip,reprint,jcp]{revtex4-1}

\usepackage{graphicx}
\usepackage{amssymb}

\newcommand{\DFA}{\mathrm{DFA}}
\newcommand{\ISI}{\mathrm{ISI}}
\newcommand{\revISI}{\mathrm{revISI}}
\newcommand{\LB}{\mathrm{LB}}
\newcommand{\SPL}{\mathrm{SPL}}
\newcommand{\GL}{\mathrm{GL2}}

\usepackage{color}

\begin{document}

\title{Assessment of interaction-strength interpolation formulas for gold and silver clusters}
\author{Sara Giarrusso}
\affiliation{Department of Theoretical Chemistry and Amsterdam Center for Multiscale Modeling, FEW, Vrije Universiteit, De Boelelaan 1083, 1081HV Amsterdam, The Netherlands}
\author{Paola Gori-Giorgi}
\affiliation{Department of Theoretical Chemistry and Amsterdam Center for Multiscale Modeling, FEW, Vrije Universiteit, De Boelelaan 1083, 1081HV Amsterdam, The Netherlands}
\author{Fabio Della Sala}
\affiliation{Institute for Microelectronics and Microsystems (CNR-IMM), Via Monteroni, Campus Unisalento, 73100 Lecce, Italy}
\affiliation{Center for Biomolecular Nanotechnologies @UNILE, Istituto Italiano di Tecnologia, Via Barsanti, I-73010
Arnesano, Italy}
\author{Eduardo Fabiano}
\affiliation{Institute for Microelectronics and Microsystems (CNR-IMM), Via Monteroni, Campus Unisalento, 73100 Lecce, Italy}
\affiliation{Center for Biomolecular Nanotechnologies @UNILE, Istituto Italiano di Tecnologia, Via Barsanti, I-73010
Arnesano, Italy}

\date{\today}

\begin{abstract}
The performance of functionals based on the idea of interpolating between the weak and the strong-interaction limits the global adiabatic-connection integrand is carefully studied for the challenging case of noble-metal clusters. Different interpolation formulas are considered and various features of this approach are analyzed. It is found that these functionals, when used as a correlation correction to Hartree-Fock, are quite robust for the description of atomization energies, while performing less well for ionization potentials. Future directions that can be envisaged from this study and a previous one on main group chemistry are discussed.
\end{abstract}

\maketitle

\section{Introduction and theoretical background}
Noble metal clusters, in particular those made of silver and gold, 
are of high interest for different areas of materials science and chemistry
as well as for technological applications.
\cite{schmidbaur90,daniel04,pyykko04,pyykko04_2,pyykko05,pyykko08,yam08,hutchins08,yamazoe14,taketoshi14,gold_book17,haruta04,takei12,ishida16,mathew14,pereiro08,novikov17,diez10,ganguly16,bernhardt05,socaciu04,vosch07}
Noble metals clusters display, in fact, peculiar properties that differ from those of
the bulk materials, due to the higher reactivity of the surface atoms. Moreover, these
properties can be often tuned by varying the size and shape of the clusters.
\cite{daniel04,taketoshi14,takei12,mathew14,khan11,tyo15,yang06,yoon07,shao10,xing06}
For these reasons, the study of the electronic properties of metal clusters is
currently a very active research field, 
\cite{hakkinen08,wang12,yi07,tanwar13,fabiano11,fabiano09,popa15,hailstone02,mckee17,duanmu15,weis02}
with many available experimental techniques.
\cite{schooss10,taylor92,johnson00,cluster_book,fielicke04,haertelt11,creaser04,lanucara14}
Nonetheless, in most cases information from theoretical calculations
is fundamental to provide a better understanding of the results and to
aid the correct interpretation of the experimental data.\cite{hakkinen08,wang12,hakkinen03,furche02,johansson08,lechtken09,weis02,blom06}

Computational studies of noble metal clusters are, however, not straightforward
\cite{reiher09} because of the small single-particle energy gap, implying a possible
multi-reference character of the electronic states, 
and due to the complex correlation effects characterizing such systems.
For these reasons, in principle an accurate description of the electronic structure can
only be achieved by high-level correlated multi-reference approaches.
\cite{mutireference_book,multiconf_chap} However, these methods
are hardly applicable for the study of clusters, due to the very high computational 
cost.
On the other hand, ``conventional'' single-reference wave-function methods
(e.g. M{\o}ller-Plesset perturbation theory \cite{mp,cremer11}, 
configuration interaction \cite{sherrill99,shavitt98}, or coupled 
cluster \cite{cc_book,cizek91}) often display important basis set and/or 
truncation errors, even for relatively small cluster sizes, which
prevent the achievement of accurate, reliable, results.
Thus, one of the most used computational tools to study noble metal clusters
is Kohn-Sham density-functional theory (DFT).\cite{ks,dft_book,dft_chap}

DFT calculations on noble metal clusters are often performed using 
a semilocal approximation for the exchange-correlation (XC) functional,
e.g. the generalized gradient approximation (GGA) \cite{functionals_chap} 
or the meta-GGA's. \cite{dellasala16} 
This is an efficient approach, 
\cite{fabiano11,popa15,mckee17,duanmu15,yang06,hakkinen03,yoon07,furche02,chen13,zanti12,baek17}
but in various cases it has also shown limited accuracy, especially in the not so
rare case when it is necessary to discriminate between isomers with 
rather similar energies (for example in the prediction of the
two- to three-dimensional crossover in gold and silver clusters
\cite{johansson08,duanmu15}).
However, unlike in the case of main group molecular calculations, the use of hybrid
functionals, which include a fraction of exact exchange, is not
able to provide a systematic improvement. Instead, it often 
leads to a worsening of the results.\cite{duanmu15,baek17}
The origin of this problem possibly traces back on the
too simplicistic idea of mixing a fixed fraction of exact 
exchange with a semilocal approximation.

In the hybrid wavefunction-DFT formalism a certain fraction $a$ of the electron-electron interaction is treated within a wave function method, while the remaining energy is captured with a semilocal functional. In a compact notation\cite{ShaTouSav-JCP-11} this can be written as 
\begin{equation}
	\label{eq_generalhybrid}
	E_0=\min_\Psi\left\{\langle\Psi|\hat{T}+a\,\hat{V}_{ee}+\hat{V}_{ne}|\Psi\rangle+\bar{E}_{\rm Hxc}^{a}[\rho_\Psi]\right\},
\end{equation}
where the complementary Hartree-exchange-correlation functional $\bar{E}_{\rm Hxc}^{a}$ depends on $\Psi$ only through its density $\rho_\Psi$. In Eq.~(\ref{eq_generalhybrid}), $\hat{T}$ is the electronic kinetic energy operator, $\hat{V}_{ee}$ the electron-electron repulsion operator and $\hat{V}_{ne}$ the external potential due to the nuclei. When the minimization over $\Psi$ in Eq.~(\ref{eq_generalhybrid}) is restricted to single Slater determinants $\Phi$, we obtain the usual hybrid functional approximation, which mixes a fraction $a$ of Hartree-Fock exchange with a semilocal functional, while using second-order peturbation theory to improve the wavefunction $\Psi$ leads to single-parameter double-hybrid functionals.\cite{ShaTouSav-JCP-11}

The XC part $E_{xc}[\rho]$ of $\bar{E}_{\rm Hxc}^{a}$  that needs to be approximated in the standard hybrid functionals formalism is usually modeled starting from the adiabatic connection formula \cite{becke93,Ern-CPL-96,pbe0,hapbe}
\begin{equation}\label{glob-xc-en}
E_{xc}[\rho]= \int _0 ^{1} W_{\lambda}[\rho] d\lambda
\end{equation}
where $\lambda$ is the interaction strength and 
${W}_{\lambda}[\rho]=\langle\Psi_\lambda[\rho]|\hat{V}_{ee}|\Psi_\lambda[\rho]\rangle - U[\rho]$
is the density-fixed linear adiabatic connection integrand, with
$\Psi_\lambda[\rho]$ being the wave function that minimizes $\hat{T}+\lambda \hat{V}_{ee}$ while yielding
the density $\rho$, and $U[\rho]$ being the Hartree energy.
Most hybrid functionals then employ a simple ansatz for the 
density-fixed linear adiabatic connection integrand, for example
\cite{pbe0,hapbe}
\begin{equation}\label{e2}
W_{\lambda}[\rho] = W_{\lambda}^{\DFA}[\rho] + \left(E_x-E_x^{\DFA}\right)(1-\lambda)^{n-1}\ ,
\end{equation}
where DFA denotes a density functional approximation (i.e. a semilocal
functional), $E_x$ denotes the Hartree-Fock exchange functional, and $n$ is a parameter.
Substituting Eq.~(\ref{e2}) into Eq.~(\ref{glob-xc-en}), yields the usual linear
mixing between the exact exchange and the density functional approximation with $a=1/n$.
However, Eq.~(\ref{e2}) is a quite arbitrary expression for $W_{\lambda}$.
It only satisfies the constraint that $W_0=E_x$ but for $\lambda\neq0$
it incorporates no exact information and it is not even 
recovering the correct weak-interaction limit behavior. 
Thus, most of the accuracy of hybrids relies on the empiricism
included into the parameter $n$ and the DFA.
This seems to work well for main-group molecular systems but not for other systems such as metal clusters considered here.

A possible non-empirical route that allows to overcome the limitations of a fixed mixing parameter
is the original idea of Seidl and coworkers\cite{SeiPerKur-PRL-00,SeiPerKur-PRA-00,seidl05} to build a model for the adiabatic-connection integrand of Eq.~(\ref{glob-xc-en}) by interpolating between the
known weak- and strong-coupling limits,\cite{gl2,GorVigSei-JCTC-09}
\begin{eqnarray}
\label{e3}
W_{\lambda\rightarrow0}[\rho] & = & W_0[\rho] + \lambda W_0'[\rho] + \cdots \\
\label{e4}
W_{\lambda\rightarrow\infty}[\rho] & = & W_{\infty}[\rho] + \frac{W_{\infty}'[\rho]}{\sqrt{\lambda}}  + \cdots\ ,
\end{eqnarray}
where
\begin{eqnarray}
W_0[\rho] & = & E_x[\rho]\ ,\\
W_0'[\rho] & = & 2 E_c^{\GL}[\rho]\ ,
\end{eqnarray}
with $E_c^{GL2}$ being the second-order G\"orling-Levy (GL) correlation
energy,\cite{gl2} whereas
$W_{\infty}[\rho]$ is the indirect part of the minimum expectation value of the 
electron-electron repulsion in a given density,\cite{Sei-PRA-99,SeiGorSav-PRA-07} and $W_{\infty}'[\rho]$ is the potential energy of 
coupled zero-point oscillations.\cite{Sei-PRA-99,GorVigSei-JCTC-09} The idea is that by using a function of $\lambda$ able to link the result from perturbation theory with the $\lambda\to\infty$ expansion of $W_\lambda[\rho]$, an approximate resummation of the perturbative series is obtained.\cite{SeiPerKur-PRA-00}

The exact $W_{\infty}[\rho]$ and $W_{\infty}'[\rho]$ are highly nonlocal density functionals\cite{SeiGorSav-PRA-07,GorVigSei-JCTC-09} 
that were approximated in the original work of Seidl and coworkers\cite{SeiPerKur-PRA-00, seidl05} by the semilocal point-charge-plus-continuum (PC) model
(see the appendix). As a result, a series of XC functionals can be derived depending on the chosen interpolating function and on whether
the $\lambda\to\infty$ expansion includes or not the order $1/\sqrt{\lambda}$: 
ISI \cite{SeiPerKur-PRL-00,SeiPerKur-PRA-00,seidl05,fabiano16} and
revISI \cite{GorVigSei-JCTC-09} also include $W_\infty'[\rho]$, while SPL \cite{SeiPerLev-PRA-99} and LB \cite{LiuBur-PRA-09} only include $W_\infty[\rho]$.
They are briefly described in the appendix.
These functionals, which are all based on an adiabatic connection integrand 
interpolation (ACII), will be generally referred to as ACII functionals.
They are non-empirical in the sense that they are approximate perturbation-theory resummations, include full exact exchange, and
describe correctly correlation in the weak-interaction limit.
Therefore, they are well-suited to try to
overcome the limitations of semilocal and hybrid DFT approaches. Their most severe problem could be the lack of size consistency for species made of different atoms, an error that is absent in the case of homogeneous clusters. Moreover, the size consistency issue is actually quite subtle\cite{GorSav-JPCS-08,Sav-CP-09} and can be corrected in many cases. 

The ACII functionals have been rarely tested on systems of interest for practical applications, with the exception of
a recent assessment of the ISI functional
for main-group chemistry.\cite{fabiano16} This investigation
has revealed interesting features of this functional and suggested
possibilities for future applications.

In this paper we move away from main group chemistry to assess different
ACII functionals for the description of the electronic properties of noble metal clusters,
made up of gold and silver. As we have 
mentioned above, these are very important systems for
materials science and chemical applications but their proper 
computational description is still a challenge. Thus, the testing 
of high-level DFT methods for this class of systems has a great practical interest.
Moreover, the application of non-empirical XC functionals,
constructed on a well defined theoretical framework, to the 
challenging problem of the simulation of electronic properties of 
noble metal clusters can help to highlight new properties and limitations
of such approaches. In fact, the next step forward could be to model the adiabatic connection integrand locally\cite{ZhoBahErn-JCP-15,VucIroSavTeaGor-JCTC-16,VucIroWagTeaGor-PCCP-17} by interpolating between the exact exchange energy density and the $\lambda\to\infty$ one, for which exact results\cite{MirSeiGor-JCTC-12} and approximations compatible with the exact exchange energy density have been recently designed.\cite{WagGor-PRA-14,BahZhoErn-JCP-16,VucGor-JPCL-17} In order to be compatible with the exact exchange energy density, these approximations are non-local and thus more expensive than the semilocal PC functionals (which suffer from the usual gauge problem that arises when we want to combine semilocal functionals with the exact exchange energy density and thus cannot be used in this framework). It has been found that the local interpolations are in general more accurate than their global counterpart.\cite{VucIroWagTeaGor-PCCP-17} Thus, the study carried out here provides also a very useful first idea of what could be achieved with these higher-level approaches.

\section{Computational details}
\label{compdet_sect}
\begin{figure}
\includegraphics[width=0.95\columnwidth]{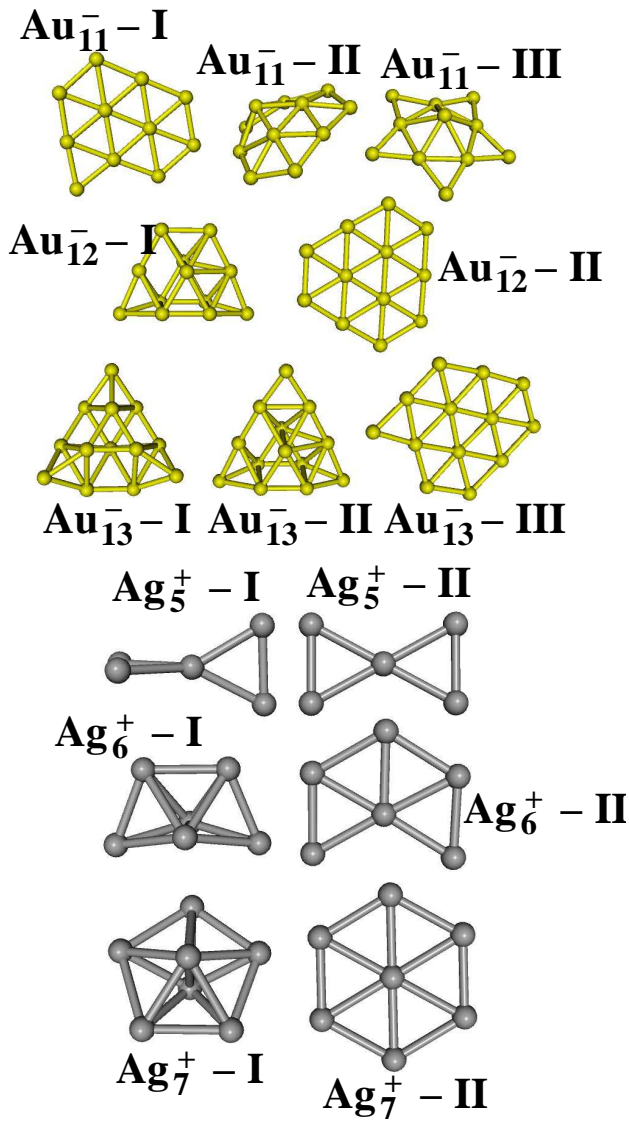}
\caption{\label{2d3d_fig}Structures of the gold and silver clusters considered for the 2D-3D dimensional crossover problem.}
\end{figure}

In this work we have tested four ACII XC functionals, which are based on
an interpolation of the density-fixed linear adiabatic connection integrand,
namely ISI,\cite{SeiPerKur-PRL-00,SeiPerKur-PRA-00,seidl05,fabiano16}
revISI,\cite{GorVigSei-JCTC-09}, SPL,\cite{SeiPerLev-PRA-99}, and LB\cite{LiuBur-PRA-09} (see the appendix
for details).
Additionally, for comparison, we have included results from the 
Perdew-Burke-Ernzerhof (PBE) \cite{pbe} and the PBE0 \cite{pbe0,pbe0_2} functionals,
which are among the most used semilocal and hybrid functionals, respectively,
as well as from the B2PLYP double hybrid functional,\cite{b2plyp} 
which also includes a fraction of second-order M\o ller-Plesset correlation energy (MP2).
We have also considered a comparison with the second- , third- and
fourth-order M\o ller-Plesset perturbation theory (MP2, MP3, MP4) \cite{mp}
results. This is because, as explained, the ACII functionals can be seen as an approximate resummation of perturbation theory, so that it is interesting to compare them with the first few lower orders. The reference results used in the assessment are specified below for each test set considered:

\begin{figure}
\includegraphics[width=0.7\columnwidth]{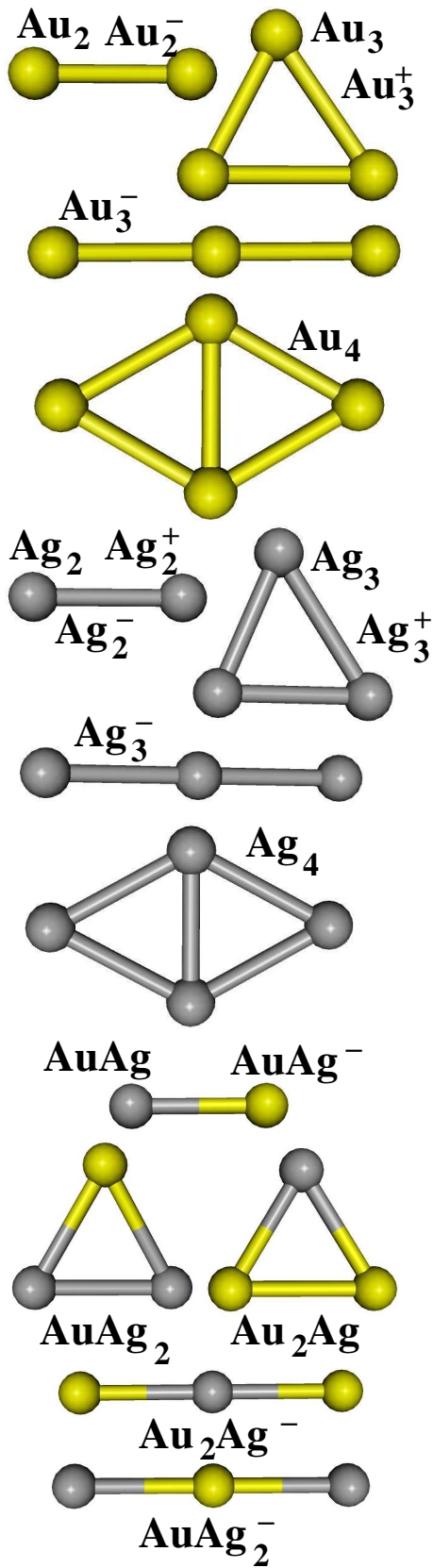}
\caption{\label{clusters_fig}Structures of the small gold, silver and binary gold-silver clusters.}
\end{figure}
\begin{itemize}
\item \textbf{Small gold clusters}. This set consists of the
Au$_2$, Au$_2^-$, Au$_3$, Au$_3^+$, Au$_3^-$, and Au$_4$ clusters.
For all these systems we have calculated the atomization energies;
for the anions as well as for Au$_3$ we have computed the ionization potential (IP)
energies. The geometries of all clusters have been taken from Ref. 
\onlinecite{fabiano11}; they are shown in Fig. \ref{clusters_fig}. 
Reference energies have been calculated at the
CCSD(T) level of theory.\cite{purvis82,pople87,scuseria88,raghavachari89}
\item \textbf{Small silver clusters}. This set includes
Ag$_2$, Ag$_2^+$, Ag$_2^-$, Ag$_3$, Ag$_3^+$, Ag$_3^-$, Ag$_4$.
As for the small gold clusters case, we have computed the
atomization energies of all the silver clusters and the IP 
of the anions as well as of Ag$_3$.
The geometries of all systems have been taken from Ref.
\onlinecite{duanmu15}; they are shown in Fig. \ref{clusters_fig}. 
Reference values for the energies have been obtained from
CCSD(T) \cite{purvis82,pople87,scuseria88,raghavachari89} calculations.
\item \textbf{Binary gold-silver clusters}. This set considers
the AuAg, AuAg$^-$, Au$_2$Ag, Au$_2$Ag$^-$, AuAg$_2$, and AuAg$_2^-$
clusters. Atomization energies have been calculated for all system,
while IPs have been computed for the anions.
Note that for the anions we considered as atomization energy the average
with respect to the two possible dissociation channels, that is
AuAg$^-$ $\rightarrow$ Au+Ag$^-$ and AuAg$^-$ $\rightarrow$ Au$^-$+Ag;
Au$_2$Ag$^-$ $\rightarrow$ Au$_2$+ Ag$^-$ and Au$_2$Ag$^-$ $\rightarrow$ Au$_2^-$+ Ag;
AuAg$_2^-$ $\rightarrow$ Au + Ag$_2^-$ and AuAg$_2^-$ $\rightarrow$ Au$^-$ + Ag$_2$.
The geometries of the binary clusters have been obtained considering the
structures reported in Ref. \onlinecite{baek17} (see Fig. \ref{clusters_fig}) 
and optimizing them at the revTPPS/def2-QZVP level of theory.\cite{revtpss,qzvp}
Reference energies have been calculated at the
CCSD(T) level of theory.\cite{purvis82,pople87,scuseria88,raghavachari89}
\item \textbf{Gold 2D-3D crossover}. This set includes the
Au$_{11}^-$, Au$_{12}^-$, and Au$_{13}^-$ clusters, that are
involved in the two- to three-dimensional crossover of gold clusters.
The geometries of all systems have been taken from Ref. \onlinecite{johansson08}
and are shown in Fig. \ref{2d3d_fig}.
\item \textbf{Silver 2D-3D crossover}. This set consists of the
Ag$_5^+$, Ag$_6^+$, and Ag$_7^+$ clusters, which are relevant to study the
two- to three-dimensional crossover of silver clusters.
Geometries have been obtained optimizing at the revTPPS/def2-QZVP level of 
theory, \cite{revtpss,qzvp} the lowest lying structures reported in Ref.
\onlinecite{duanmu15}. The structures are reported in Fig. \ref{2d3d_fig}.
\end{itemize}

All the required calculations have been performed with the
TURBOMOLE program package, \cite{turbo1,turbo2} 
employing, unless otherwise stated, 
the aug-cc-pwCVQZ-PP basis set 
\cite{hill12} and a Stuttgart-Koeln MCDHF 60-electron effective core potential.
\cite{figgen05}
The calculations concerning the ISI, revISI, SPL, and LB functionals 
have been performed in a post-self-consistent-field (post-SCF) fashion,
using Hartree-Fock orbitals. This choice is consistent with the
results of Ref.~\onlinecite{fabiano16}, where it has been found that the ISI functional yields much better results when
used as a correlation correction for the HF energy.
The PBE and PBE0 calculations have been performed using a full SCF procedure;
B2PLYP calculations have been carried out as described in Ref. \onlinecite{b2plyp},
considering a SCF tretment of the exchange and semilocal correlation part and
adding the second-order MP2 correlation fraction as a post-SCF correction.

\section{Results}
\label{res_sect}
In this section we analyze the performance of the ACII XC funcionals for
the description of the electronic properties of gold, silver and mixed Au/Ag
clusters. The results are compared to those obtained from other 
approaches, such as semilocal and hybrid DFT as well as wave-function perturbation theory.
\subsection{Total Energies}
To start our investigation we consider, in table \ref{tab_xc},
the errors on total energies computed with different methods
with respect to the CCSD(T) reference values.
\begin{table*}
\caption{\label{tab_xc}Errors on total energies (eV/atom) of small gold, silver, and binary clusters. For each set of clusters the mean absolute error (MAE) is reported. In the bottom part of the table we report also the statistics for the overall set (mean error (ME), MAE, and standard deviation).}
\begin{ruledtabular}
\begin{tabular}{lrrrrrrrrrr}
 & PBE & PBE0 &	B2PLYP & ISI & revISI & SPL & LB & MP2 & MP3 & MP4  \\
\hline
Au  &  -4.93  &  -3.75  &  -1.73  &  2.84  &  2.93  &  2.65  &  1.97  &  -0.33  &  0.98  &  -0.27    \\ 
Au+  &  -4.58  &  -3.73  &  -1.64  &  2.63  &  2.70  &  2.50  &  1.89  &  -0.10  &  0.68  &  -0.15      \\ 
Au-  &  -4.94  &  -3.47  &  -1.65  &  3.26  &  3.38  &  3.02  &  2.25  &  -0.35  &  1.44  &  -0.41      \\ 
Au2  &  -4.96  &  -3.66  &  -1.69  &  2.89  &  2.99  &  2.67  &  1.94  &  -0.53  &  1.22  &  -0.43      \\ 
Au2-  &  -4.97  &  -3.58  &  -1.66  &  3.06  &  3.17  &  2.83  &  2.08  &  -0.46  &  1.40  &  -0.44      \\ 
Au3  &  -4.97  &  -3.65  &  -1.67  &  2.89  &  3.00  &  2.67  &  1.94  &  -0.54  &  1.29  &  -0.48      \\ 
Au3+  &  -4.90  &  -3.68  &  -1.66  &  2.80  &  2.90  &  2.59  &  1.88  &  -0.50  &  1.15  &  -0.43      \\ 
Au3-  &  -4.94  &  -3.54  &  -1.66  &  2.98  &  3.10  &  2.73  &  1.96  &  -0.65  &  1.40  &  -0.55      \\ 
Au4  &  -4.96  &  -3.62  &  -1.67  &  2.86  &  2.97  &  2.63  &  1.87  &  -0.66  &  1.33  &  -0.54      \\ 
ME  &  -4.91  &  -3.63  &  -1.67  &  2.91  &  3.01  &  2.70  &  1.97  &  -0.46  &  1.21  &  -0.41      \\ 
MAE  &  4.91  &  3.63  &  1.67  &  2.91  &  3.01  &  2.70  &  1.97  &  0.46  &  1.21  &  0.41      \\ 
  &    &    &    &    &    &    &    &    &    &        \\ 
Ag  &  -0.89  &  -0.36  &  0.21  &  3.24  &  3.39  &  2.91  &  2.18  &  -0.26  &  1.04  &  -0.27      \\ 
Ag+  &  -0.44  &  -0.25  &  0.38  &  2.99  &  3.13  &  2.71  &  2.04  &  -0.19  &  0.76  &  -0.20      \\ 
Ag-  &  -0.99  &  -0.21  &  0.24  &  3.68  &  3.85  &  3.31  &  2.53  &  -0.08  &  1.39  &  -0.32      \\ 
Ag2  &  -0.96  &  -0.30  &  0.22  &  3.32  &  3.49  &  2.97  &  2.19  &  -0.37  &  1.18  &  -0.38      \\ 
Ag2+  &  -0.79  &  -0.34  &  0.28  &  3.14  &  3.28  &  2.82  &  2.11  &  -0.26  &  0.95  &  -0.27      \\ 
Ag2-  &  -1.00  &  -0.27  &  0.24  &  3.50  &  3.67  &  3.14  &  2.37  &  -0.22  &  1.34  &  -0.33      \\ 
Ag3  &  -0.95  &  -0.30  &  0.25  &  3.32  &  3.49  &  2.96  &  2.19  &  -0.39  &  1.23  &  -0.40      \\ 
Ag3+  &  -0.85  &  -0.30  &  0.27  &  3.21  &  3.37  &  2.86  &  2.11  &  -0.41  &  1.10  &  -0.38      \\ 
Ag3-  &  -0.97  &  -0.23  &  0.25  &  3.45  &  3.63  &  3.07  &  2.27  &  -0.38  &  1.32  &  -0.43      \\ 
Ag4  &  -0.95  &  -0.27  &  0.24  &  3.29  &  3.47  &  2.91  &  2.12  &  -0.51  &  1.24  &  -0.45      \\ 
ME  &  -0.88  &  -0.28  &  0.26  &  3.31  &  3.48  &  2.97  &  2.21  &  -0.31  &  1.15  &  -0.34      \\ 
MAE  &  0.88  &  0.28  &  0.26  &  3.31  &  3.48  &  2.97  &  2.21  &  0.31  &  1.15  &  0.34      \\ 
  &    &    &    &    &    &    &    &    &    &        \\ 
AuAg  &  -2.93  &  -1.95  &  -0.72  &  3.10  &  3.24  &  2.81  &  2.06  &  -0.47  &  1.20  &  -0.41      \\ 
AuAg-  &  -2.97  &  -1.91  &  -0.70  &  3.27  &  3.41  &  2.97  &  2.20  &  -0.37  &  1.36  &  -0.39      \\ 
Au$_2$Ag  &  -3.58  &  -2.49  &  -1.00  &  3.03  &  3.16  &  2.76  &  2.01  &  -0.51  &  1.28  &  -0.46      \\ 
Au$_2$Ag$^-$  &  -3.57  &  -2.40  &  -1.00  &  3.14  &  3.28  &  2.84  &  2.06  &  -0.57  &  1.36  &  -0.50      \\ 
AuAg$_2$  &  -2.26  &  -1.38  &  -0.37  &  3.18  &  3.33  &  2.86  &  2.10  &  -0.46  &  1.26  &  -0.43      \\ 
AuAg$_2$$^-$  &  -2.32  &  -1.34  &  -0.40  &  3.30  &  3.46  &  2.96  &  2.18  &  -0.46  &  1.36  &  -0.47      \\ 
ME  &  -2.94  &  -1.91  &  -0.70  &  3.17  &  3.31  &  2.87  &  2.10  &  -0.47  &  1.30  &  -0.44      \\ 
MAE  &  2.94  &  1.91  &  0.70  &  3.17  &  3.31  &  2.87  &  2.10  &  0.47  &  1.30  &  0.44      \\ 
  &    &    &    &    &    &    &    &    &    &        \\ 
\multicolumn{11}{c}{Overall statistics} \\ 
ME  &  -2.82  &  -1.88  &  -0.66  &  3.13  &  3.27  &  2.85  &  2.10  &  -0.40  &  1.21  &  -0.39      \\ 
MAE  &  2.82  &  1.88  &  0.87  &  3.13  &  3.27  &  2.85  &  2.10  &  0.40  &  1.21  &  0.39      \\ 
Std.Dev.  &  1.81  &  1.51  &  0.87  &  0.24  &  0.27  &  0.18  &  0.16  &  0.15  &  0.20  &  0.10      \\ 
\end{tabular}
\end{ruledtabular}
\end{table*}
Although this quantity is usually not of much interest in practical applications 
(where energy differences are usually considered), the analysis of the 
errors on total energies will be useful to understand the performances of the
different functionals for more practical properties such as atomization or ionization energies.

Inspection of the data shows that the ACII functionals do not perform very well for the total energy.
In fact, they yield the highest mean absolute errors (MAEs), being even slightly worse than the
semilocal PBE approach and giving definitely larger errors with respect to perturbation theory
(MP2, MP3, and MP4) and to the double hybrid B2PLYP functional. Among the ACII functionals,
the SPL and especially the LB approach perform systematically better than ISI and revISI.
Thus LB yields errors which are often 30\% smaller than ISI, even though they are 
still usually larger than those of the other non-ACII methods.
On the other hand, considering the standard deviation of the errors (last line of Table \ref{tab_xc})
we note that the ACII results display a quite small dispersion around the average 
(with LB and SPL again slightly better than ISI and revISI). This is related to the fact that
the ACII functionals all give a quite systematic underestimation (in magnitude) of the energy of all systems.
In contrast, PBE, PBE0, and partly B2PLYP give larger values of the standard deviation.
This depends on the fact that these methods describe quite accurately some systems (e.g. Ag clusters),
which are the ones that effectively contribute to produce a quite low MAE, but
they give significantly larger errors for other systems. This behavior is a signature of the
too simplicistic nature of these functionals, which cannot capture equally well the physics of all systems.

The observed standard deviations suggest that, when energy differences are considered,
the ACII functionals can benefit from a cancellation of the systematic error, such that
rather accurate energy differences can be obtained. 
We must remark also that the standard deviation values reported in Table \ref{tab_xc} 
allow only a partial understanding of the problem because they are obtained from all the
data but, depending on the property of interest, some energy differences may be more relevant 
than others, e.g. for atomization energies the difference between
a cluster energy and the energy of the composing atoms is the most relevant. Thus, for example
MP methods all yield quite low standard deviations, but a closer look at the results shows that
the errors for atoms are quite different than those for the clusters (much more different than
for ACII methods); hence, we can expect that, despite a quite good MAE and a small standard deviation,
MP2, MP3, and MP4 atomization energies can display a limited accuracy.
A more detailed analysis of the relationship between the data reported in Table \ref{tab_xc}
and some relevant energy difference properties will be given in Section \ref{discuss_sect}.

\subsection{Atomization and Ionization energies}
A first example of an important energy difference is the atomization energy.
The atomization energy values calculated for the sets of gold, silver, and binary clusters with
all the methods are reported in Table \ref{tab1}.
\begin{table*}
\caption{\label{tab1}Atomization energies (eV) of small gold, silver, and binary clusters. Note that for anionic binary clusters the average between the two possible dissociation paths has been considered (see Section \ref{compdet_sect}). For each set of clusters the mean error (ME), the mean absolute error (MAE), the mean absolute relative error (MARE), and the standard deviation are reported. In the bottom part of the table we report also the statistics for the overall set.}
\begin{ruledtabular}
\begin{tabular}{lrrrrrrrrrrc}
 & PBE & PBE0 &	B2PLYP & ISI & revISI & SPL & LB & MP2 & MP3 & MP4 & CCSD(T) \\
\hline
Au$_2$  &  2.33  &  2.08  &  2.20  &  2.17  &  2.14  &  2.24  &  2.33  &  2.67  &  1.79  &  2.60  &  2.27  \\ 
Au$_2^-$  &  1.97  &  1.83  &  1.83  &  1.86  &  1.84  &  1.90  &  1.95  &  2.14  &  1.51  &  2.09  &  1.89  \\ 
Au$_3$  &  3.57  &  3.14  &  3.26  &  3.28  &  3.23  &  3.39  &  3.54  &  4.08  &  2.51  &  4.07  &  3.45  \\ 
Au$_3^+$  &  6.06  &  5.60  &  5.67  &  5.71  &  5.66  &  5.82  &  5.97  &  6.54  &  4.98  &  6.38  &  5.79  \\ 
Au$_3^-$  &  4.90  &  4.52  &  4.73  &  4.87  &  4.81  &  5.00  &  5.17  &  5.80  &  4.05  &  5.57  &  4.87  \\ 
Au$_4$  &  6.18  &  5.51  &  5.81  &  5.95  &  5.85  &  6.14  &  6.40  &  7.37  &  4.60  &  7.10  &  6.03  \\ 
ME  &  0.12  &  -0.27  &  -0.14  &  -0.08  &  -0.13  &  0.03  &  0.18  &  0.71  &  -0.81  &  0.58  &    \\ 
MAE  &  0.12  &  0.27  &  0.14  &  0.08  &  0.13  &  0.06  &  0.18  &  0.71  &  0.81  &  0.58  &    \\ 
MARE  &  3\%  &  7\%  &  3\%  &  2\%  &  4\%  &  1\%  &  4\%  &  17\%  &  21\%  &  14\%  &    \\ 
Std.Dev.  &  0.08  &  0.16  &  0.06  &  0.06  &  0.07  &  0.08  &  0.13  &  0.39  &  0.37  &  0.30  &    \\ 
  &    &    &    &    &    &    &    &    &    &    &    \\ 
Ag$_2$  &  1.82  &  1.59  &  1.69  &  1.53  &  1.50  &  1.59  &  1.66  &  1.93  &  1.41  &  1.91  &  1.70  \\ 
Ag$_2^+$  &  1.85  &  1.69  &  1.64  &  1.58  &  1.57  &  1.59  &  1.61  &  1.70  &  1.51  &  1.67  &  1.62  \\ 
Ag$_2^-$  &  1.53  &  1.39  &  1.37  &  1.32  &  1.31  &  1.35  &  1.38  &  1.51  &  1.15  &  1.48  &  1.41  \\ 
Ag$_3$  &  2.73  &  2.37  &  2.45  &  2.31  &  2.27  &  2.41  &  2.52  &  2.94  &  2.00  &  2.94  &  2.56  \\ 
Ag$_3^+$  &  4.84  &  4.45  &  4.50  &  4.36  &  4.32  &  4.47  &  4.59  &  5.03  &  4.05  &  4.90  &  4.52  \\ 
Ag$_3^-$  &  3.70  &  3.32  &  3.49  &  3.38  &  3.32  &  3.50  &  3.63  &  4.12  &  3.06  &  4.01  &  3.57  \\ 
Ag$_4$  &  4.80  &  4.24  &  4.47  &  4.39  &  4.30  &  4.58  &  4.80  &  5.59  &  3.78  &  5.28  &  4.59  \\ 
ME  &  0.19  &  -0.13  &  -0.05  &  -0.16  &  -0.20  &  -0.07  &  0.03  &  0.41  &  -0.43  &  0.32  &    \\ 
MAE  &  0.19  &  0.15  &  0.06  &  0.16  &  0.20  &  0.07  &  0.06  &  0.41  &  0.43  &  0.32  &    \\ 
MARE  &  7\%  &  5\%  &  2\%  &  6\%  &  7\%  &  3\%  &  2\%  &  13\%  &  15\%  &  10\%  &    \\ 
Std.Dev.  &  0.07  &  0.14  &  0.06  &  0.07  &  0.09  &  0.05  &  0.09  &  0.32  &  0.23  &  0.22  &    \\ 
  &    &    &    &    &    &    &    &    &    &    &    \\ 
AuAg  &  2.22  &  1.97  &  2.11  &  2.05  &  2.02  &  2.13  &  2.21  &  2.53  &  1.80  &  2.46  &  2.18  \\ 
AuAg$^-$  &  1.83  &  1.69  &  1.71  &  1.74  &  1.72  &  1.78  &  1.83  &  2.00  &  1.47  &  1.92  &  1.77  \\ 
Au$_2$Ag  &  3.65  &  3.26  &  3.42  &  3.47  &  3.41  &  3.59  &  3.73  &  4.28  &  2.80  &  4.22  &  3.65  \\ 
Au$_2$Ag$^-$  &  4.96  &  4.58  &  4.84  &  4.97  &  4.91  &  5.12  &  5.28  &  5.90  &  4.36  &  5.63  &  5.04  \\ 
AuAg$_2$  &  3.33  &  2.94  &  3.08  &  3.06  &  3.00  &  3.17  &  3.30  &  3.80  &  2.56  &  3.75  &  3.28  \\ 
AuAg$_2^-$  &  3.83  &  3.40  &  3.58  &  3.48  &  3.42  &  3.60  &  3.74  &  4.25  &  3.00  &  4.14  &  3.63  \\ 
ME  &  0.04  &  -0.29  &  -0.14  &  -0.13  &  -0.18  &  -0.03  &  0.09  &  0.53  &  -0.60  &  0.43  &    \\ 
MAE  &  0.07  &  0.29  &  0.14  &  0.13  &  0.18  &  0.06  &  0.09  &  0.53  &  0.60  &  0.43  &    \\ 
MARE  &  2\%  &  9\%  &  4\%  &  4\%  &  6\%  &  2\%  &  3\%  &  16\%  &  19\%  &  13\%  &    \\ 
Std.Dev.  &  0.09  &  0.14  &  0.08  &  0.07  &  0.08  &  0.06  &  0.08  &  0.23  &  0.21  &  0.18  &    \\ 
  &    &    &    &    &    &    &    &    &    &    &    \\
\multicolumn{12}{c}{Overall statistics} \\ 
ME  &  0.12  &  -0.22  &  -0.11  &  -0.12  &  -0.17  &  -0.02  &  0.10  &  0.54  &  -0.60  &  0.44  &    \\ 
MAE  &  0.13  &  0.23  &  0.11  &  0.12  &  0.17  &  0.06  &  0.11  &  0.54  &  0.60  &  0.44  &    \\ 
MARE  &  4\%  &  7\%  &  3\%  &  4\%  &  6\%  &  2\%  &  3\%  &  15\%  &  18\%  &  12\%  &    \\ 
Std.Dev.  &  0.10  &  0.16  &  0.08  &  0.07  &  0.08  &  0.07  &  0.11  &  0.33  &  0.31  &  0.25  &    \\ 
\end{tabular}
\end{ruledtabular}
\end{table*}
Observing the data it appears that, as anticipated, for atomization energies
the ACII functionals work fairly well. In particular, SPL and LB, yield 
mean absolute relative errors (MAREs) of about 2-3\% for all kinds of
clusters, being competitive with the B2PLYP functional. The ISI and revISI 
functionals perform slighlty worse, displaying a systematic underbinding and
giving overall MAREs of 4\% and 6\%, respectively. Moreover, unlike for SPL and LB, 
non-negligible differences exist in the description of the different materials with
gold clusters described better than silver ones. Overall the
ISI and revISI functionals show a comparable performance as PBE and better than PBE0.
Finally, the MP results show a quite poor performance, exhibiting MAREs ranging form 10\%
to 20\%. 
In addition, we can note that MP2 results are closer to MP4 results than MP3 ones 
not only from a quantitative point of view but also qualitatively (MP2 and MP4 always 
overbind while MP3 always consistently underbinds). This is a clear indication of
the difficult convergence of the perturbative series for the metal clusters electronic 
properties. 

In Table \ref{tab2}, we report the computed ionization potential energies, which are 
other important energy differences to consider for metal clusters.
\begin{table*}
\caption{\label{tab2}Ionization potentials (eV) of small gold, silver, and binary clusters. For each set of clusters the mean error (ME), the mean absolute error (MAE), the mean absolute relative error (MARE), and the standard deviation are reported. In the bottom part of the table we report also the statistics for the overall set.}
\begin{ruledtabular}
\begin{tabular}{lrrrrrrrrrrc}
 & PBE & PBE0 &	B2PLYP & ISI & revISI & SPL & LB & MP2 & MP3 & MP4 & CCSD(T) \\
\hline 
Au  &  9.54  &  9.22  &  9.29  &  9.00  &  8.97  &  9.05  &  9.13  &  9.42  &  8.91  &  9.32  &  9.20  \\ 
Au$^-$  &  2.30  &  2.00  &  2.21  &  1.86  &  1.84  &  1.92  &  2.01  &  2.31  &  1.82  &  2.42  &  2.29  \\ 
Au$_2^-$  &  1.94  &  1.75  &  1.84  &  1.56  &  1.55  &  1.58  &  1.62  &  1.78  &  1.53  &  1.91  &  1.91  \\ 
Au$_3$  &  7.05  &  6.76  &  6.89  &  6.57  &  6.55  &  6.62  &  6.69  &  6.97  &  6.44  &  7.01  &  6.86  \\ 
Au$_3^-$  &  3.63  &  3.38  &  3.67  &  3.45  &  3.41  &  3.53  &  3.63  &  4.03  &  3.36  &  3.92  &  3.70  \\ 
ME  &  0.10  &  -0.17  &  -0.01  &  -0.30  &  -0.33  &  -0.25  &  -0.17  &  0.11  &  -0.38  &  0.12  &    \\ 
MAE  &  0.13  &  0.18  &  0.06  &  0.30  &  0.33  &  0.25  &  0.17  &  0.16  &  0.38  &  0.12  &    \\ 
MARE  &  2\%  &  6\%  &  2\%  &  10\%  &  11\%  &  9\%  &  6\%  &  4\%  &  12\%  &  3\%  &    \\ 
Std.Dev.  &  0.17  &  0.14  &  0.07  &  0.08  &  0.08  &  0.09  &  0.11  &  0.18  &  0.07  &  0.08  &    \\ 
  &    &    &    &    &    &    &    &    &    &    &    \\ 
Ag  &  8.04  &  7.70  &  7.76  &  7.35  &  7.33  &  7.40  &  7.45  &  7.67  &  7.31  &  7.66  &  7.59  \\ 
Ag$^-$  &  1.40  &  1.15  &  1.28  &  0.86  &  0.85  &  0.90  &  0.95  &  1.13  &  0.95  &  1.35  &  1.31  \\ 
Ag$_2$  &  8.02  &  7.60  &  7.80  &  7.30  &  7.26  &  7.40  &  7.50  &  7.90  &  7.21  &  7.90  &  7.68  \\ 
Ag$_2^-$  &  1.11  &  0.96  &  0.97  &  0.66  &  0.65  &  0.66  &  0.67  &  0.72  &  0.69  &  0.92  &  1.01  \\ 
Ag$_3$  &  5.93  &  5.63  &  5.71  &  5.30  &  5.28  &  5.34  &  5.39  &  5.58  &  5.26  &  5.70  &  5.64  \\ 
Ag$_3^-$  &  2.38  &  2.10  &  2.32  &  1.93  &  1.90  &  1.99  &  2.06  &  2.31  &  2.02  &  2.42  &  2.31  \\ 
ME  &  0.22  &  -0.06  &  0.05  &  -0.36  &  -0.38  &  -0.31  &  -0.25  &  -0.04  &  -0.35  &  0.07  &    \\ 
MAE  &  0.22  &  0.10  &  0.07  &  0.36  &  0.38  &  0.31  &  0.25  &  0.14  &  0.35  &  0.10  &    \\ 
MARE  &  6\%  &  5\%  &  2\%  &  17\%  &  17\%  &  15\%  &  13\%  &  8\%  &  15\%  &  4\%  &    \\ 
Std.Dev.  &  0.16  &  0.11  &  0.08  &  0.07  &  0.07  &  0.07  &  0.09  &  0.18  &  0.07  &  0.10  &    \\ 
  &    &    &    &    &    &    &    &    &    &    &    \\ 
AuAg$^-$  &  1.46  &  1.30  &  1.35  &  1.05  &  1.04  &  1.07  &  1.09  &  1.19  &  1.07  &  1.34  &  1.39  \\ 
Au$_2$Ag$^-$  &  3.16  &  2.90  &  3.17  &  2.87  &  2.84  &  2.94  &  3.03  &  3.34  &  2.95  &  3.30  &  3.18  \\ 
AuAg$_2^-$  &  2.35  &  2.04  &  2.24  &  1.79  &  1.76  &  1.84  &  1.91  &  2.17  &  1.83  &  2.28  &  2.15  \\ 
ME  &  0.09  &  -0.16  &  0.01  &  -0.34  &  -0.36  &  -0.29  &  -0.23  &  -0.01  &  -0.29  &  0.07  &    \\ 
MAE  &  0.10  &  0.16  &  0.05  &  0.34  &  0.36  &  0.29  &  0.23  &  0.13  &  0.29  &  0.10  &    \\ 
MARE  &  5\%  &  7\%  &  3\%  &  17\%  &  18\%  &  15\%  &  12\%  &  7\%  &  15\%  &  4\%  &    \\ 
Std.Dev.  &  0.11  &  0.11  &  0.07  &  0.03  &  0.03  &  0.04  &  0.07  &  0.18  &  0.05  &  0.10  &    \\ 
  &    &    &    &    &    &    &    &    &    &    &    \\ 
\multicolumn{12}{c}{Overall statistics} \\ 
ME  &  0.15  &  -0.12  &  0.02  &  -0.33  &  -0.36  &  -0.28  &  -0.22  &  0.02  &  -0.35  &  0.09  &    \\ 
MAE  &  0.16  &  0.14  &  0.06  &  0.33  &  0.36  &  0.28  &  0.22  &  0.15  &  0.35  &  0.11  &    \\ 
MARE  &  4\%  &  6\%  &  2\%  &  14\%  &  15\%  &  13\%  &  11\%  &  6\%  &  14\%  &  4\%  &    \\ 
Std.Dev.  &  0.16  &  0.12  &  0.08  &  0.07  &  0.07  &  0.07  &  0.09  &  0.18  &  0.07  &  0.09  &    \\
\end{tabular}
\end{ruledtabular}
\end{table*}
In this case the ACII functionals perform rather poorly, being the worst methods, if we exclude MP3.
As in the case of atomization energies, SPL and LB (especially the latter) show a slightly better
performance than ISI and revISI. Nevertheless, the results are definitely worst than for B2PLYP, PBE
and even PBE0. A rationalization of this failure will be given is section \ref{discuss_sect}.

\subsection{2D-3D crossover}
To conclude this section, we consider the problem of the two- to three-dimensional (2D-3D) 
crossover of anionic gold clusters and cationic silver clusters.
Different studies have indicated that for anionic gold clusters the dimensional crossover
occurs between Au$_{11}^-$ (2D) and Au$_{13}^-$ (3D), with the 2D and 3D Au$_{12}^-$
structures being almost isoenergetic.\cite{furche02,johansson08}
On the other hand, for cationic silver clusters it has been suggested that the
dimensional transition occurs already for Ag$_5^+$, which has a 2D structure 
with a slightly lower energy than the 3D one, while Ag$_6^+$ and
Ag$_7^+$ display lowest energy 3D structures.\cite{weis02,vandertol17}
Anyway, this is a quite difficult problem because experimentally it is not trivial to distinguish
clusters of the same size but different dimensionality. A computational support is thus
required.\cite{weis02,furche02,johansson08,vandertol17,kruckeberg96,hakkinen03}
However, to describe correctly the energy ordering of several noble metal clusters
with very similar energies is a hard task for any computational method.\cite{johansson08,fabiano11,mantina09,bloc}
For this reason, this is a very interesting problem from the computational point of view.

In Table \ref{tab3}, we report the energies calculated for the 
anionic gold clusters and cationic silver clusters
relevant for the 2D-3D transition.
\begin{table*}
\caption{\label{tab3}Relative energies (eV) with respect to conformer I (see Computational details) of 2D and 3D anionic gold clusters and cationic silver clusters. For the gold clusters the data include the correction terms reported in Table IV of Ref. \onlinecite{johansson08}.}
\begin{ruledtabular}
\begin{tabular}{lcrrrrrrrrr}
 & & PBE & PBE0 & BLOC & B2PLYP & ISI & revISI & SPL & LB & MP2 \\
\hline
Au$_{11}^-$-I & 2D &  --  &  --  &  --  &  --  &  --  &  --  &  --  &  --  &  --  \\
Au$_{11}^-$-II & 3D &  0.217  &  0.224  &  0.206  &  0.147  &  0.083  &  0.090  &  0.070  &  0.054  &  -0.006  \\
Au$_{11}^-$-III & 3D &  0.270  &  0.179  &  0.354  &  0.254  &  0.265  &  0.251  &  0.302  &  0.344  &  0.499  \\
Au$_{12}^-$-I & 3D &  --  &  --  &  --  &  --  &  --  &  --  &  --  &  --  &  --  \\
Au$_{12}^-$-II & 2D &  -0.450  &  -0.340  &  0.008  &  -0.144  &  0.710  &  0.669  &  0.789  &  0.882  &  1.228  \\
Au$_{13}^-$-I & 3D &  --  &  --  &  --  &  --  &  --  &  --  &  --  &  --  &  --  \\
Au$_{13}^-$-II & 3D &  -0.027  &  -0.032  &  0.037  &  -0.024  &  0.497  &  0.495  &  0.499  &  0.527  &  0.618  \\
Au$_{13}^-$-III & 2D &  -0.111  &  0.056  &  0.386  &  0.248  &  0.802  &  0.894  &  0.917  &  0.824  &  1.069  \\
  &    &    &    &    &    &    &    &    &    \\
Ag$_5^+$-I & 3D & --  & -- & -- & -- & -- & -- & -- & -- & -- \\
Ag$_5^+$-II & 2D &  0.021  &  0.025  &  0.024  &  0.020  &  0.021  &  0.020  &  0.018  &  0.017  &  0.013  \\
Ag$_6^+$-I & 3D & --  & -- & -- & -- & -- & -- & -- & -- & -- \\
Ag$_6^+$-II & 2D &  -0.005  &  0.055  &  0.280  &  0.007  &  0.220  &  0.211  &  0.241  &  0.265  &  0.348  \\
Ag$_7^+$-I & 3D & --  & -- & -- & -- & -- & -- & -- & -- & -- \\
Ag$_7^+$-II  & 2D &  -0.099  &  --  &  0.303  &  -0.059  &  0.286  &  0.270  &  0.318  &  0.352  &  0.474  \\
\end{tabular}
\end{ruledtabular}
\end{table*}
The Table shows, for comparison, also the results obtained with the
BLOC meta-GGA functional \cite{bloc,constantin12,constantin13}, which
is expected to be one of the most accurate approaches for this kind of problems.
Observing the data, one can immediately note that the PBE, PBE0 and even B2PLYP methods
 are not reliable for the dimensional crossover of noble metal clusters.
In fact, PBE always favors 2D structures, whereas PBE0 predicts the 2D-3D transition
at a too large cluster dimension for gold, Au$_{13}^-$ (although 
the 3D geometry with lowest energy is not the same as the one 
we find with BLOC and all ACII functionals),
 and for silver the energies of the 2D and 3D clusters differ 
 slightly for both $n=6$ and $n=7$, not evidencing a clear transition 
 at the expected cluster size. 
A similar behaviour is found for the
B2PLYP functional, which was instead one of the best for the atomization
energies and IPs of small clusters.
The ACII functionals overall perform all quite similarly,
predicting for all clusters the expected ordering and
 agreeing  well with BLOC results for the cationic Ag clusters but 
tending to favor 3D structures in the anionic Au clusters.
We note that this behavior is somehow inherited from the MP2
method, which however performs much worse than any of the ACII functionals
considered here.

\section{Discussion and Analysys of the results}
\label{discuss_sect}

In the previous Section we saw that the ACII functionals perform rather well for the
calculation of atomization energies of noble metal clusters.
As mentioned above, a good rationalization of the observed results can be obtained in terms
of the energy errors that the different methods display for the total energies of atoms and
of the clusters. These have been reported in Table \ref{tab_xc}. 
\subsection{Energy differences}
\begin{figure}
\includegraphics[width=1.0\columnwidth]{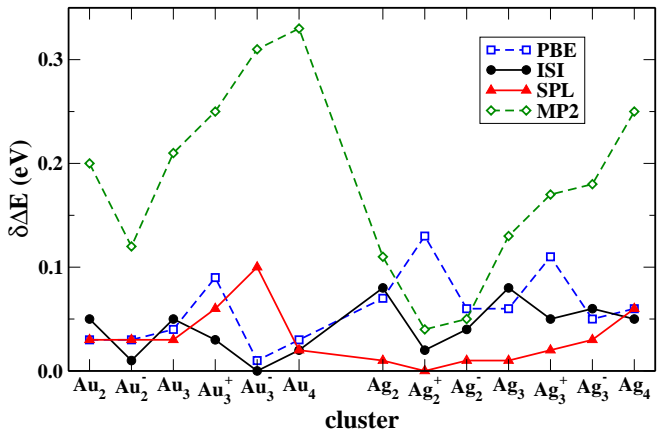}
\caption{\label{dd_fig}Difference in the total energy error between a cluster and its constituent atoms (see Eq. (\ref{e5})).}
\end{figure}
For a better visualization here we additionaly plot, in Fig. \ref{dd_fig}, the
quantity
\begin{eqnarray}
\label{e5}
\delta\Delta E & = & \Delta E(M_nM_m^-M_l^+) - \\
\nonumber
& - & \sum_{n}\Delta E(M_n)  - \sum_{m}\Delta E(M_n^-)- \sum_{l}\Delta E(M_n^+)\ ,
\end{eqnarray}
where $\Delta E$ are the total energy errors (the $\Delta E$ per atom are 
reported in Table \ref{tab_xc}),
$M$=Au or Ag, and $n,m,l$ are integers such that $M_nM_m^-M_l^+$ corresponds to 
a given cluster (e.g. for Au$_3^+$ we have $M$=Au, $n=2$, $m=0$, and $l=1$).
This quantity provides a measure of how different is the energy error made on a
given cluster from that of its constituent atoms.
Inspection of the plots shows that the smaller $\delta\Delta E$ values
are yielded by the ISI and SPL (revISI and LB, not reported, give similar results).
These functionals are also among the best performers for the atomization energies.
On the other hand, for PBE we observe that the $\delta\Delta E$ is
small for gold clusters, with the exception of Au$_3^+$, while for
silver clusters is larger. Indeed, looking to Table \ref{tab1} we can find
that PBE performs well for gold clusters, with the exception of Au$_3^+$
that yields an error of 0.27 eV (more than twice larger than the MAE), while
it performs less well for silver clusters.
Finally, for MP2 the values of $\delta\Delta E$ are generally very large.
Thus, despite MP2 is on average quite accurate in the description of
the total energies (see Table \ref{tab_xc}) it fails to produce
accurate atomization energies because of accumulation of the errors.

\begin{figure}
\includegraphics[width=1.0\columnwidth]{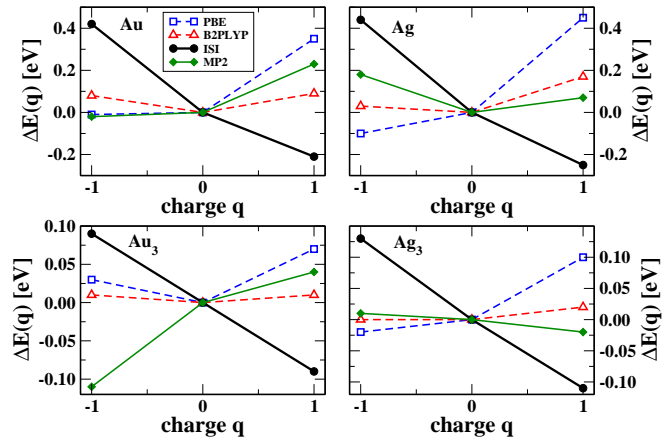}
\caption{\label{fig_charge}Variation of the energy error with the total charge of the system (Au top left, Au$_3$ bottom left, Ag top right, Ag$_3$ bottom right). The values are scaled to the neutral system value (see Eq. (\ref{e6})).}
\end{figure}

A similar analysis, can be
made to comment the results of the ionization potential calculations
(reported in Table \ref{tab2}). However, in this case the difference to 
consider is between the neutral and the charged species.
Then, a different behavior is observed.
In fact, while for most of the considered methods the total energy error is
not much different between a neutral and a charged species of the
same cluster, for the ACII functionals we always observe an increase
of the error with the charge.
This situation is schematized in Fig. \ref{fig_charge},
where we plot, for several examples, the quantity
\begin{equation}\label{e6}
\Delta(q) = \Delta E(A^q) - \Delta E(A^0)\ ,
\end{equation}
with $A$ being any of the systems under investigation and
$q=-1,0,1$.
The observed trend may trace back to a different ability of ACII
functionals to describe the high- and low-density regimes.
As a consequence, the ACII functionals are generally the worst performers
for the calculation of ionization potentials, while PBE
and especially B2PLYP perform well thanks to the more homogeous description
of the differently charged species.

This analysis shows that, although the quality of the total energies produced by
a functional is a key element to understand the performance of the
functional, the basic property to observe is not the
 quality of the absolute energies, but rather the variance of the errors. 
Furthermore, the contrasting behaviors we have observed for the description of the
atomization energies and of the ionization potentials
 highlights the subtleties inherent to such calculations.
In particular, the accuracy of the ACII functionals has been shown to be not 
much dependent on the investigated material (Au or Ag) nor on the system's
size but to be quite sensitive to the charge state of the computed system.
The first feature is a positive one. This is related, as we saw,
to the computation of atomization energies, but even more importantly it
indicates that the idea beyond the 
construction of the ACII functionals is in general quite robust such that
the functionals, although not very accurate in absolute terms (see Table \ref{tab_xc}) 
are well transferabe to systems of different size and composition.
This is not a trivial results since, as we documented, other methods 
(e.g. PBE and PBE0, but even MP4) do not share this property.
On the contrary, the dependence of the ACII functionals on the
charge state of the system indicates a clear limitation of such approaches.
They are in fact unable to describe with similar accuracy systems with
qualitatively different charge distributions. As a consequence, the ionization
potential calculations are problematic for ACII functionals.\\
Note however that, because accurate experimental data are not available
for all the systems, our assessment of the performances of the ACII functionals on small clusters,
see Fig. \ref{clusters_fig}, and Tables \ref{tab_xc}, \ref{tab1}, and \ref{tab2},
is carried out w.r.t. CCSD(T) values. This allows a more direct and sensible comparison of
the results, whereas the comparison with experimental data would require
the consideration of further effects such as thermal/vibronic ones as well as spin-orbit coupling.\cite{fabiano11,johansson08}
Of course CCSD(T) results cannot be considered ``exact" for metal clusters. Nevertheless,
an accurate comparison with available experimental data from literature shows that, 
for atoms (regarding ionization energies)  \cite{LooBeaSim-PRA-99, HaaWan-JPC-03}
and neutral dimers and trimers (regarding both ionization and atomization energies),  
\cite{BalFen-CPL-89, HubHer-MSMS-79, BeuDem-JCP-93}
 CCSD(T) yields results within 0.04 eV  
from the experimental ones.
While for the charged dimers and trimers (regarding both ionization and 
atomization energies), CCSD(T) results are within 0.2 eV \cite{SpaErv-JCP-99, LeeKim-JCPB-03, HaaWan-JPC-03}
from the experimental ones. This larger discrepancy may be 
partly ascribable to a diminished accuracy of the CCSD(T) calculation
\emph{per se} in these cases, but it may also be  
possibly due to the rather large error bars associated to the measures
on the experimental side and on the increased importance of
correcting terms on the computational side.

\subsection{AC curves: gold dimer showcase}

\begin{figure}
\includegraphics[width=1.0\columnwidth]{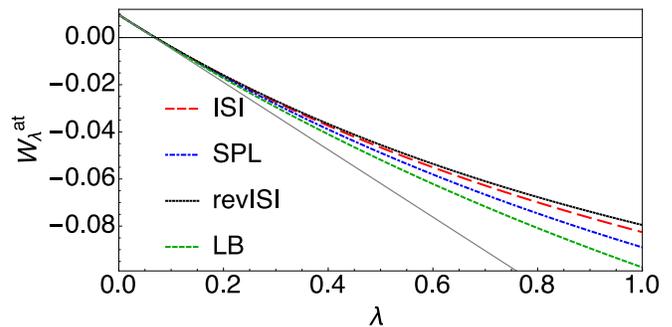}
\caption{\label{ac_fig}Atomization adiabatic connection integrands (see Eq. (\ref{e7})) corresponding to ISI, revISI, SPL, and LB for the Au$_2$ case; the thick curve in gray corresponds to the linear expansion for the atomization adiabatic connection integrand (Eq. (\ref{e8}))
}
\end{figure}

To rationalize the origin of the limitations of the ACII functionals 
as well as to understand in depth the differences and the similarities between 
the different interpolation formulas it would be necessary to inspect
in some detail the shape of the density-fixed linear adiabatic 
connection integrand defining ISI, revISI, SPL, and LB.
However, contrary to small atoms and molecules (see, e.g., Refs.~\onlinecite{ColSav-JCP-99,TeaCorHel-JCP-09,TeaCorHel-JCP-10}), for noble metal clusters there exists no reference
adiabatic connection integrands to compare to. Thus, such a detailed
analysis is not really possible.
Nevertheless, some useful hints can be obtained by a semi-qualitative comparison
of the various adiabatic connection curves.
As an example, in Fig.~\ref{ac_fig} we report, for the Au$_2$ case (the other systems studied here have very similar features), 
the atomization adiabatic connection integrand, defined as
\begin{equation}\label{e7}
W^{at}_\lambda(\mathrm{Au}_2) = W_\lambda(\mathrm{Au}_2) - 2W_\lambda(\mathrm{Au})\ ,
\end{equation}
for ISI, revISI, SPL, and LB. The integrated value 
(between 0 and 1) of this
quantity corresponds to the XC atomization energy calculated 
with a given ACII functional.
For discussion we have plotted also the weak interacting limit expansion truncated at linear order in $\lambda$
for the atomization adiabatic connection integrand, which is defined as
\begin{equation}\label{e8}
W^{at}_{\lambda,LE}(\mathrm{Au}_2) = W_{\lambda,LE} (\mathrm{Au}_2) - 2 W_{\lambda,LE} (\mathrm{Au})  ,
\end{equation}
where the linear expansion (LE) of the AC integrand for a species X is
 $W_{\lambda,LE} (\mathrm{X}) = E_x(\mathrm{X}) +
  2\lambda E_c^{\GL}(\mathrm{X})$ in agreement with Eq.~(\ref{e3}) 
  and in the case of HF orbitals $E_c^{\GL}(\mathrm{X}) =E_c^{\mathrm{MP2}}(\mathrm{X})$.
Because of the weak-interacting limit constraint, 
all the curves plotted in the figure share the same $\lambda=0$ value,
which corresponds to the Hartree-Fock exchange atomization energy, as well as 
the same slope at this point.
The curves remain very similar up to $\lambda\approx 0.2$, which is
not strictly dictated by the weak-interacting limit constraint but rather
by a possible lack of flexibility in the interpolation formulas.
For values of $\lambda \gtrsim 0.2$, the curves associated to the various
functionals start to differ, due to the different ways they approach the
$W_\infty$ value for $\lambda=\infty$. Note that in this case
ISI and revISI are further constrained to recover the $W_\infty'$ slope,
whereas SPL and LB do not have this constraint.
The interpolation towards the strong-interaction limit is therefore the
main feature differentiating the various ACII functionals, even in the range $0\leq\lambda\leq1$.
In general, revISI is the slowest to approach the
asymptotic $W_\infty$ value, whereas LB is the fastest. So the former 
will usually yield the smaller XC energies, whereas the latter will produce the
larger XC energies (in magnitude).
In fact, turning to the Au$_2$ example reported in Fig.~\ref{ac_fig}, the inspection 
of the plot shows that revISI is indeed the slowest to move towards the
asymptotic $W_\infty^{at}$ value (for Au$_2$ $W_\infty^{at}=-0.239$).
Consequently, in Table \ref{tab1} it yields the smallest atomization energy (it
underestimates the Au$_2$ atomization energy by 0.13 eV). On the
opposite, LB is the fastest to move towards the asymptotic $W_\infty^{at}$
value, thus it gives the larger atomization energy (overestimating
it by 0.06 eV).
In this specific case, the SPL functional, which behaves almost intermediately
between revISI and LB, yields a very accurate value of the atomization energy,
underestimating it by only 0.03 eV.

Thus, we have seen that there are two main features that
can determine the performance of an ACII functional. The first one is
surely the behavior towards the strong-coupling limit, which is
able to influence the shape of the adiabatic connection integrand curve
for $\lambda\gtrapprox0.2/0.3$. This behavior is indeed modeled differently
by the various functionals examined in this work, but it appears that
none of them can really capture the correct behavior in the
range of interest $0.3\le\lambda\le1$. This is possibly due
to the fact that information on the $\lambda=\infty$ point
is not sufficient to guide correctly the interpolation at the quite small $\lambda$
values of interest for the calculation of XC energies.
A second factor that is relevant for the functionals' performance
is the small $\lambda$ behavior. At very small $\lambda$ values this is
determined by Eq. (\ref{e3}), but for larger values of the coupling constant
(at least for $0.1\leq\lambda\leq0.2$) the shape of the curve 
should depart from the slope given by $E_c^{\GL}$ in order to
correctly describe the  higher-order correlation effect.
Instead, we have observed that all the ACII functionals provide the same behavior
up to $\lambda\approx0.2$. This indicates that the interpolation
formulas have not enough flexibility to differentiate from the
asymptotic behavior imposed at $\lambda=0$.

\subsection{Role of the reference orbitals}
\label{sec_reforb}
The ACII functionals are orbital-dependent nonlinear functionals,
thus they are usually employed to compute the XC energy in a post-SCF fashion
(as we did in this work). 
Then, the results depend on the choice of the orbitals used for the calculation.
Recent work \cite{fabiano16} has evidenced that ISI results for main-group chemistry 
are much improved when Hartree-Fock orbitals are used. 
This has been basically traced back to the characteristics of the Hartree-Fock
single-particle energy gap (which determines the magnitute of $E_c^{\GL}$
and thus the weak-interaction behavior of the curves).

\begin{figure}
\includegraphics[width=1.0\columnwidth]{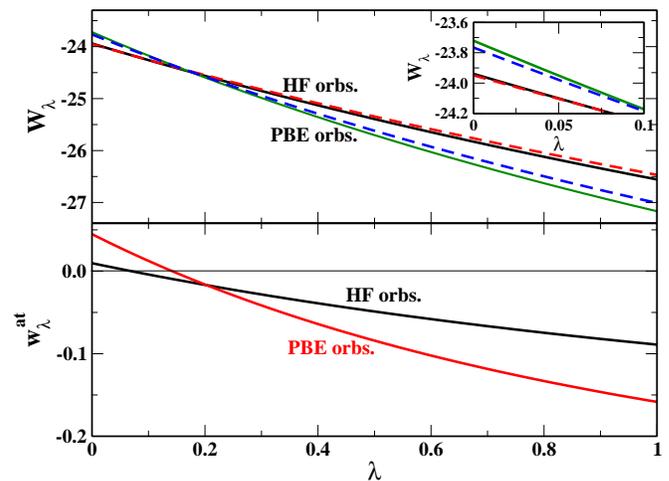}
\caption{\label{spl_ac_fig}Top: Adiabatic connection integrands computed with the SPL formula [Eq. (\ref{spl_eq})] for Au$_2$ (solid line) and Au (dashed line) using Hartree-Fock and PBE orbitals; the Au curve is multiplied by a factor of 2; the inset shows the weak-interaction part of the curves. Bottom: Atomization adiabatic connection integrands (see Eq. (\ref{e7})) computed with the SPL formula for the Au$_2$ case.}
\end{figure}

For gold and silver clusters, after some test calculations, we found a similar result
for all the ACII formulas considered.
For this reason, all the results reported in Section \ref{res_sect} are
based on Hartree-Fock orbitals.
To clarify this aspect, we have reported in Fig. \ref{spl_ac_fig}
both the bare and the atomization adiabatic connection integrands 
computed with the SPL formula (similar results are obtained for 
the other formulas) for Au$_2$ and Au using either Hartree-Fock
and PBE orbitals. 
It can be seen that the adiabatic connection curve of Au$_2$,
obtained from Hartree-Fock orbitals, is very similar to twice the
Au curve. Hence, the atomization adiabatic connection integrand
is rather flat, yielding (correctly) a moderate atomization XC energy.
This behavior depends partly on the fact that in Hartree-Fock calculations
Au$_2$ has almost twice the exchange energy of Au but, primarly,
it traces back to the fact that the Au$_2$ 
MP2 correlation energy is almost perfeclty two times larger than the
Au one (which in turn depends on the fact that the two systems have
very close single-particle energy gaps -- 7.604 eV and 7.707 eV, respectively --
and on the size-extensivity of the MP2 method).
Thus, the adiabatic connection integrands for Au$_2$ and twice the Au
have almost identical slopes at $\lambda=0$ and similar behaviors
for $\lambda\leq 1$.
Instead, when PBE orbitals are used, larger differences between the
Au$_2$ and twice the Au curves can be noted.
These originate only partially from the fact that, in the case of PBE orbitals,
the exact exchange contributions of Au$_2$ and twice Au 
are not much similar (they differ by 0.045 eV).
Mostly they depend on the rather different GL2 correlation
energies for the systems ($E_c^\GL(\mathrm{Au}_2)-2E_c^{GL2}(\mathrm{Au})=-0.173 eV$),
which in turn trace back to the fact that the single particle energy gaps
computed for Au$_2$ and Au are very different: 2.014 eV and 0.718 eV, respectively.
Consequently, the atomization adiabatic connection integrand curve
calculated with PBE orbitals is steeper than the Hartree-Fock-based one
and therefore it yields significantly larger atomization XC energies.
This results in a strong tendency of PBE-based ACII functionals to overbind 
the noble metal clusters.

\subsection{Further analysis of the ACII's formulas}
\label{sec-discInterp}
We have seen in Sec.~\ref{res_sect} that SPL and LB formulas show 
overall better performances than ISI and revISI. 
As mentioned, the main difference between the two groups is that 
the former use a three-parameters interpolation formula
 while the latter make use of a fourth ingredient
 from the $\lambda \rightarrow \infty$ limit, i.e. the zero-point oscillation term $W_\infty'[\rho]$. The revISI formula also recovers the exact expansion at large $\lambda$ to higher orders.\cite{GorVigSei-JCTC-09}
However, we have to keep in mind that the ingredients coming from 
the strong interaction limit are not computed exactly, 
but approximated with the semilocal PC model. 
Comparison with the exact $W_\infty[\rho]$ and $W_\infty'[\rho]$
for light atoms\cite{SeiGorSav-PRA-07,GorVigSei-JCTC-09} suggests that the PC approximation of the $W_\infty[\rho]$ term 
is more accurate than the one for $W_\infty'[\rho]$. 
Moreover, the parameters appearing in the PC model for $W_\infty[\rho]$ are 
all determined by the electrostatics of the PC cell, while in the case of $W_\infty'[\rho]$
 the gradient expansion does not give a physical result, 
 and one of the parameters has to be fixed in other ways, 
 for example by making the model exact for the He atom.\cite{GorVigSei-JCTC-09} 

Another important point to consider is that, as explained in Sec.~\ref{sec_reforb}, we are using the ACII functionals with Hartree-Fock orbitals, which means that they are used as a correlation functional for the Hartree-Fock energy. In other words, the ACII correlation functionals are used here as an approximate resummation of the M{\o}ller-Plesset perturbation series: they recover the exact MP2 at weak coupling, and perform much better than MP3 and MP4 for atomization energies (see Table \ref{tab1}). Thus, a first question that needs to be addressed is whether the PC model used here to compute the infinite coupling strength functionals is accurate also for the Hartree-Fock adiabatic connection, in which the $\lambda$-dependent hamiltonian reads
\begin{equation}
	\label{eq_adiaHF}
	\hat{H}^{\lambda}=\hat{T}+\hat{V}_{\rm HF}+\lambda\,(\hat{V}_{ee}-\hat{V}_{\rm HF}),
\end{equation}
with $\hat{V}_{\rm HF}$ the Hartree-Fock non local potential operator. When $\lambda\to\infty$, the problem defined by $\hat{H}^{\lambda}$ of Eq.~(\ref{eq_adiaHF}) is not the same as the one of the density-fixed adiabatic connection arising in DFT. The results of this study may suggest that the PC model can provide a decent approximation of the leading $\lambda\to\infty$ term in the HF adiabatic connection integrand, at least when dealing with isoelectronic energy differences. A careful study of the problem is the object of on-going work.

Keeping in mind that the information from $W_\infty'[\rho]$ is less accurate (and maybe less relevant in the HF context), it can be interesting to consider a variant of ISI and revISI, in which
we replace $W_\infty'^{\mathrm{PC}}[\rho]$ with the curvature at $\lambda=0$ (obtained from MP3) as input ingredient. In this way,
the modified AC integrand expressions recover
the first three terms of Eq.~(\ref{e3}) for small $\lambda$, 
and only the first term of Eq.~(\ref{e4}) for large $\lambda$. However, 
the resulting XC approximations show several drawbacks.
In fact, the results for atomization energies are significantly worse
than for the original ISI and revISI functionals (the MAEs are 0.72 and 
0.65 eV for Au clusters, 0.40 and  0.37 eV for Ag clusters and 
0.55  and 0.51 eV for binary clusters), despite they are close to MP2
ones and better than MP3 ones. More importantly,
the modified ISI and revISI formulas, 
with the input ingredients for the Au and Ag clusters,
result in adiabatic connection integrands that become imaginary at some
$\lambda > 1$, with revISI breaking down at much larger $\lambda$ values than
ISI. This fact might be ascribed to the oscillatory behavior of the MP series, which gives a curvature that is too large, or to the lack of flexibility of the revISI and ISI formulas. This is further illustrated in the appendix.

\section{Conclusions and perspectives}
We have assessed the performance of functionals based on the idea of interpolating between the weak and the strong-interaction limits the global adiabatic-connection integrand (ACII functionals) for noble-metal clusters, analyzing and rationalizing different features of this approach. The study presented here extends a previous preliminary assessment on main group chemistry,\cite{fabiano16} and explores different interpolation formulas.

We have found that the ACII functionals, although not spectacularly accurate, are quite robust for the description of atomization energies, as their performance tends to be the same for different species and different cluster sizes, which is a positive feature. We should also stress that this good performance is achieved by using 100\% of Hartree-Fock exchange, and thus avoiding to rely on error cancellation between exchange and correlation. Rather, as clearly shown in fig.~\ref{dd_fig}, this is achieved by performing in a very similar way for the description of a cluster and its constituent atoms.
On the other hand, the ACII functionals are found to be inaccurate for ionization energies, as they are not capable to describe differently charged states of the same system with the same accuracy, as shown in fig.~\ref{fig_charge}.

As in the case of main-group chemistry,\cite{fabiano16} we have found that the ACII functionals perform much better when used with Hartree-Fock orbitals, which means that they are used as a correlation functional for the Hartree-Fock energy. In other words, the ACII correlation functionals are used here as an approximate resummation of the M{\o}ller-Plesset perturbation series: they recover the exact MP2 at weak coupling, and perform much better than MP3 and MP4 for atomization energies (see Table \ref{tab1}). Thus, a first question that needs to be addressed is whether the PC model used here to compute the infinite coupling strength functionals is accurate also for the Hartree-Fock adiabatic connection of Eq.~(\ref{eq_adiaHF}), which is the object of a current investigation. The results of this study and of Ref.~\onlinecite{fabiano16} suggest that the PC model can provide a decent approximation of the $\lambda\to\infty$ HF adiabatic connection integrand, at least when dealing with isoelectronic energy differences. 

Another promising future direction is the development of ACII functionals in
which the interpolation is done in each point of space, on energy
densities.\cite{ZhoBahErn-JCP-15,VucIroSavTeaGor-JCTC-16,
  VucIroWagTeaGor-PCCP-17} These local interpolations are more amenable to
construct size-consistent approximations, but need energy densities all
defined in the same gauge (the one of the electrostatic potential of the
exchange-correlation hole seems so far to be the most suitable for this
purpose \cite{VucLevGor-JCP-17}). In this framework, the simple PC model, which performs globally quite well, does not provide accurate approximations pointwise,\cite{VucIroWagTeaGor-PCCP-17} and needs to be replaced with models based on integrals of the spherically averaged density,\cite{WagGor-PRA-14,BahZhoErn-JCP-16} which, in turn, needs a careful implementation, which is the focus of on-going efforts.\cite{BahZhoErn-JCP-16} Finally, recent models for $\lambda=1$ could be also used in this framework, \cite{VucGor-JPCL-17} both locally and globally.

\begin{acknowledgments}
Financial support was provided by the European Research Council under H2020/ERC Consolidator Grant corr-DFT [Grant Number 648932]. 
We thank TURBOMOLE GmbH for providing the TURBOMOLE program package.
\end{acknowledgments}

\appendix

\section{Adiabatic connection integrand interpolation formulas}

Several interpolation formulas have been developed to
recover the weak- and strong-coupling limit behaviours
of Eqs. (\ref{e3}) and (\ref{e4}).
For the sake of simplicity, we will not specify in the following that the expressions of the AC integrand as well as of the XC correlation energy are (explicit or implicit) functionals of the density as well as each of their fundamental ingredients, $W_0, W_0', W_0'', W_\infty, $ and $W_\infty '$.

\noindent{\bf  Interaction Strength Interpolation (ISI) formula}\cite{SeiPerKur-PRL-00,SeiPerKur-PRA-00,seidl05,fabiano16}\\
\begin{equation}
W_\lambda^\ISI  = W_\infty + \frac{X }{\sqrt{1+\lambda Y }+Z }\ ,
\end{equation}
with
\begin{eqnarray}\label{Y}
&&X=\frac{xy^2}{z^2}\; ,\; Y=\frac{x^2y^2}{z^4}\; , \; Z=\frac{xy^2}{z^3}-1\ ;\\
&& x=-2 W_0' ,\; y=W_\infty'\; , \; z=W_0-W_\infty\ .
\end{eqnarray}
After integration in Eq. (\ref{glob-xc-en}), it gives
\begin{equation}
E_{xc}^\ISI = W_\infty + \frac{2X}{Y}\left[\sqrt{1+Y}-1-Z\ln\left(\frac{\sqrt{1+Y}+Z}{1+Z}\right)\right]\ .
\end{equation}

\noindent{\bf Revised ISI (revISI) formula}\cite{GorVigSei-JCTC-09}\\ 
\begin{equation}
W_\lambda^\revISI  = W_\infty  + \frac{b  \left( 2 + c  \lambda + 2 d  \sqrt{1 + c \lambda}\right)}{2 \sqrt{1 + c \lambda} \left( d  + \sqrt{1 + c \lambda}\right) ^2} ,
\end{equation}
where
\begin{eqnarray}\label{c}
\nonumber
b & = &-\frac{4 W_0' (W_\infty')  ^{2}}{\left(W_0-W_\infty\right)^2}\; ,\; c=\frac{2 (W_0' W_\infty ')^2}{\left(W_0-W_\infty\right)^4}\; ,\; \\
d & = & -1-\frac{4 W_0' (W_\infty ')  ^{2}}{\left(W_0-W_\infty\right)^3} \ .
\end{eqnarray}
The corresponding XC functional is
\begin{equation}
E_{xc}^\revISI = W_\infty + \frac{b}{\sqrt{1+c}+d}\ .
\end{equation}

\noindent{\bf Seidl-Perdew-Levy (SPL) formula}\cite{SeiPerLev-PRA-99}\\
\begin{equation}\label{spl_eq}
W_\lambda^\SPL  = W_\infty  +\frac{W_0 -W_\infty }{\sqrt{1+2\lambda \chi }}\ ,
\end{equation}
with
\begin{equation}
\chi = \frac{W_0'}{W_\infty-W_0}\ .
\end{equation}
The SPL XC functional reads
\begin{equation}
E_{xc}^\SPL = \left(W_0-W_\infty\right)\left[\frac{\sqrt{1+2\chi}-1-\chi}{\chi}\right] + W_0\ .
\end{equation}
Note that this functional does not make use on information on $W_\infty'$.

\noindent{\bf Liu-Burke (LB) formula}\cite{LiuBur-PRA-09}\\
\begin{equation}
W_\lambda^\LB  = W_\infty  + \beta (y  + y^4 )\ , 
\end{equation}
where 
\begin{equation}
y = \frac{1}{\sqrt{1+\gamma\lambda}}\; , \; \beta=\frac{W_0-W_\infty}{2}\; , \; \gamma=\frac{4 W_0'}{5(W_\infty-W_0)}\ .
\end{equation}
Using Eq.~(\ref{glob-xc-en}), the LB XC functional is found to be
\begin{equation}
E_{xc}^\LB = 2\beta\left[\frac{1}{\gamma}\left(\sqrt{1+c}-\frac{1+c/2}{1+c}\right)-1\right]\ .
\end{equation}
Also the LB functional does not use information on $W_\infty'$.
\\
\\
\noindent {\bf Point-Charge-plus-continuum (PC) model}\\
In all cases, the highly non-local functionals $W_\infty$ and $W_\infty'$ (when used)
are approximated by the semilocal PC model \cite{SeiPerKur-PRA-00}
\begin{eqnarray}
W_\infty  & \approx & W_{\infty}^\mathrm{PC}  = \int\left[A\rho(\mathbf{r}^{4/3}) + B\frac{|\nabla\rho(\mathbf{r})|^2}{\rho(\mathbf{r})^{4/3}}\right] d\mathbf{r}\\
W_\infty'  & \approx & {W_{\infty}'}^\mathrm{PC}  = \int\left[C\rho(\mathbf{r}^{3/2}) + D\frac{|\nabla\rho(\mathbf{r})|^2}{\rho(\mathbf{r})^{7/6}}\right] d\mathbf{r}\ ,
\end{eqnarray}
where $A=-9(4\pi/3)^{1/3}/10$, $B=3[3/(4\pi)]^{1/3}/350$, $C=\sqrt{3\pi}/2$, $D=-0.028957$;
note that other slightly different values are possible for the $D$ parameter.\cite{GorVigSei-JCTC-09}

\subsection{ISI and revISI with the exact curvature}
The ISI and revISI formulas have four parameters that need to be fixed by four equations. In the standard forms (see above) the four equations are obtained by imposing that $ W_{\lambda}^{\rm ISI}$ recovers the first two terms of the weak-interacting limit expansion, Eq. (\ref{e3}), and the first two terms in the strongly-interacting limit expansion, Eq. (\ref{e4}) for large $\lambda$. 
For the first time we have explored an alternative choice that is to constrain ISI and revISI to recover the first three terms of Eq. (\ref{e3}) for small $\lambda$, and only the first term of Eq. (\ref{e4}). 

The structure of the interpolation formula is thus formally the same, 
but the parameters are given by
\begin{eqnarray}
X =& -2 (W_0 - W_\infty) + \frac{W_0''}{(W_0')^2}(W_0 - W_\infty)^2 \nonumber,  \\ 
Y = & -2 \frac{W_0''}{W_0'} + \frac{4 W_0'}{(W_0 - W_\infty)}, \nonumber  \\
Z = & -3 + \frac{W_0''}{(W_0')^2}(W_0 - W_\infty) ;
\end{eqnarray}
for ISI, and 
\begin{eqnarray}
b =& -2 (W_0 - W_\infty) + \frac{4 W_0'' (W_0 - W_\infty)^2}{3 (W_0')^2}  \nonumber,  \\ 
c = &  -\frac{4 W_0''}{3 W_0'} + \frac{2 W_0'}{(W_0 - W_\infty)} , \nonumber  \\
d = &-3 + \frac{4 W_0'' (W_0 - W_\infty)}{3 (W_0')^2} ;
\end{eqnarray}
 for revISI. 

However, as discussed in Sec.~\ref{sec-discInterp}, while in the standard ISI and revISI interpolation formulas the parameters, $Y[\rho]$ and $c[\rho]$, which appear under square root, are given by the sum of squared quantities (see Eqs.\ref{Y}, and \ref{c}), in these modified versions this is not true and they can become negative. In the cases studied here both parameters turn out to be always negative and smaller than one,  meaning that there is, for each species, a critical lambda, $\lambda_c$, always larger than one, after which the function takes imaginary values. 
In particular we found an average $\bar{\lambda}_c^{ISI} \approx 4$ with values spanning from $2.5$ to $5.7$, and an average $\bar{\lambda}_c^{revISI} \approx 180 $ with values spanning from $6$ to over $3 \times 10^{3}$. As a general trend we thus see that the modified revISI appears to be more robust than the modified ISI in the sense that it becomes imaginary at significantly larger $\lambda$ values.

\bibliography{isi_gold,biblioPaola,biblio_add,biblio_add2,biblio_spec,biblio1}

\end{document}